\pdfoutput=1
\documentclass[cits,11pt,a4paper]{article}
\usepackage{jinstpub}

\usepackage[utf8]{inputenc}
\usepackage{amsmath}
\usepackage{amsfonts}
\usepackage{amssymb}
\usepackage{graphicx}
\usepackage{epsfig}
\usepackage{tabularx}
\usepackage{lineno}
\usepackage[section]{placeins}

\title{Radio Frequency and DC High Voltage Breakdown of High Pressure Helium, Argon, and Xenon}
\collaboration{The NEXT Collaboration}
\author[1,a]{K.~Woodruff,\note[a]{Corresponding author.}}
\author[1]{J.~Baeza-Rubio,}
\author[1]{D.~Huerta,}
\author[1]{B.J.P.~Jones,}
\author[1]{A.D.~McDonald,}
\author[1]{L.~Norman,}
\author[1]{D.R.~Nygren,}
\author[10]{C.~Adams,}
\author[17]{V.~\'Alvarez,}
\author[6]{L.~Arazi,}
\author[18]{I.J.~Arnquist,}
\author[4]{C.D.R~Azevedo,}
\author[19]{K.~Bailey,}
\author[20]{F.~Ballester,}
\author[17]{J.M.~Benlloch-Rodr\'{i}guez,}
\author[12]{F.I.G.M.~Borges,}
\author[1]{N.K.~Byrnes,}
\author[17]{S.~C\'arcel,}
\author[17]{J.V.~Carri\'on,}
\author[21]{S.~Cebri\'an,}
\author[18]{E.~Church,}
\author[12]{C.A.N.~Conde,}
\author[10]{T.~Contreras,}
\author[1]{A.A.~Denisenko,}
\author[2,14]{G.~D\'iaz,}
\author[17]{J.~D\'iaz,}
\author[5]{M.~Diesburg,}
\author[12]{J.~Escada,}
\author[20]{R.~Esteve,}
\author[17]{R.~Felkai,}
\author[11]{A.F.M.~Fernandes,}
\author[11]{L.M.P.~Fernandes,}
\author[14,8]{P.~Ferrario,}
\author[4]{A.L.~Ferreira,}
\author[1]{F.W.~Foss Jr.,}
\author[11]{E.D.C.~Freitas,}
\author[14]{J.~Generowicz,}
\author[7]{A.~Goldschmidt,}
\author[2]{D.~Gonz\'alez-D\'iaz,}
\author[14,8,b]{J.J.~G\'omez-Cadenas,\note[b]{NEXT Co-spokesperson.}}
\author[10]{S.~Ghosh,}
\author[10]{R.~Guenette,}
\author[9]{R.M.~Guti\'errez,}
\author[10]{J.~Haefner,}
\author[19]{K.~Hafidi,}
\author[3]{J.~Hauptman,}
\author[11]{C.A.O.~Henriques,}
\author[2]{J.A.~Hernando~Morata,}
\author[14,17]{P.~Herrero,}
\author[20]{V.~Herrero,}
\author[19]{S.~Johnston,}
\author[17]{M.~Kekic,}
\author[16]{L.~Labarga,}
\author[1]{A.~Laing,}
\author[5]{P.~Lebrun,}
\author[17]{N.~L\'opez-March,}
\author[9]{M.~Losada,}
\author[11]{R.D.P.~Mano,}
\author[10]{J.~Mart\'in-Albo,}
\author[14]{A.~Mart\'inez,}
\author[2,17]{G.~Mart\'inez-Lema,}
\author[14]{F.~Monrabal,}
\author[11]{C.M.B.~Monteiro,}
\author[20]{F.J.~Mora,}
\author[14]{J.~Mu\~noz~Vidal,}
\author[17]{P.~Novella,}
\author[1,c]{D.R.~Nygren,\note[c]{NEXT Co-spokesperson.}}
\author[17]{B.~Palmeiro,}
\author[5]{A.~Para,}
\author[17,d]{J.~P\'erez,\note[d]{Now at Laboratorio Subterr\'aneo de Canfranc, Spain.}}
\author[17]{M.~Querol,}
\author[17]{J.~Renner,}
\author[19]{J.~Repond,}
\author[19]{S.~Riordan,}
\author[15]{L.~Ripoll,}
\author[9]{Y.~Rodriguez Garcia,}
\author[20]{J.~Rodr\'iguez,}
\author[1]{L.~Rogers,}
\author[14]{B.~Romeo,}
\author[17]{C.~Romo-Luque,}
\author[12]{F.P.~Santos,}
\author[11]{J.M.F. dos~Santos,}
\author[6]{A.~Sim\'on,}
\author[13,e]{C.~Sofka,\note[e]{Now at University of Texas at Austin, USA.}}
\author[17]{M.~Sorel,}
\author[13]{T.~Stiegler,}
\author[1]{P.~Thapa,}
\author[20]{J.F.~Toledo,}
\author[14]{J.~Torrent,}
\author[17]{A.~Us\'on,}
\author[4]{J.F.C.A.~Veloso,}
\author[13]{R.~Webb,}
\author[6]{R.~Weiss-Babai,}
\author[13,f]{J.T.~White,\note[f]{Deceased.}}
\author[17]{N.~Yahlali}
\emailAdd{katherine.woodruff@uta.edu}
\affiliation[1]{
Department of Physics, University of Texas at Arlington, Arlington, TX 76019, USA}
\affiliation[2]{
Instituto Gallego de F\'isica de Altas Energ\'ias, Univ.\ de Santiago de Compostela, Campus sur, R\'ua Xos\'e Mar\'ia Su\'arez N\'u\~nez, s/n, Santiago de Compostela, E-15782, Spain}
\affiliation[3]{
Department of Physics and Astronomy, Iowa State University, 12 Physics Hall, Ames, IA 50011-3160, USA}
\affiliation[4]{
Institute of Nanostructures, Nanomodelling and Nanofabrication (i3N), Universidade de Aveiro, Campus de Santiago, Aveiro, 3810-193, Portugal}
\affiliation[5]{
Fermi National Accelerator Laboratory, Batavia, IL 60510, USA}
\affiliation[6]{
Nuclear Engineering Unit, Faculty of Engineering Sciences, Ben-Gurion University of the Negev, P.O.B. 653, Beer-Sheva, 8410501, Israel}
\affiliation[7]{
Lawrence Berkeley National Laboratory (LBNL), 1 Cyclotron Road, Berkeley, CA 94720, USA}
\affiliation[8]{
Ikerbasque, Basque Foundation for Science, Bilbao, E-48013, Spain}
\affiliation[9]{
Centro de Investigaci\'on en Ciencias B\'asicas y Aplicadas, Universidad Antonio Nari\~no, Sede Circunvalar, Carretera 3 Este No.\ 47 A-15, Bogot\'a, Colombia}
\affiliation[10]{
Department of Physics, Harvard University, Cambridge, MA 02138, USA}
\affiliation[11]{
LIBPhys, Physics Department, University of Coimbra, Rua Larga, Coimbra, 3004-516, Portugal}
\affiliation[12]{
LIP, Department of Physics, University of Coimbra, Coimbra, 3004-516, Portugal}
\affiliation[13]{
Department of Physics and Astronomy, Texas A\&M University, College Station, TX 77843-4242, USA}
\affiliation[14]{
Donostia International Physics Center (DIPC), Paseo Manuel Lardizabal, 4, Donostia-San Sebastian, E-20018, Spain}
\affiliation[15]{
Escola Polit\`ecnica Superior, Universitat de Girona, Av.~Montilivi, s/n, Girona, E-17071, Spain}
\affiliation[16]{
Departamento de F\'isica Te\'orica, Universidad Aut\'onoma de Madrid, Campus de Cantoblanco, Madrid, E-28049, Spain}
\affiliation[17]{
Instituto de F\'isica Corpuscular (IFIC), CSIC \& Universitat de Val\`encia, Calle Catedr\'atico Jos\'e Beltr\'an, 2, Paterna, E-46980, Spain}
\affiliation[18]{
Pacific Northwest National Laboratory (PNNL), Richland, WA 99352, USA}
\affiliation[19]{
Argonne National Laboratory, Argonne, IL 60439, USA}
\affiliation[20]{
Instituto de Instrumentaci\'on para Imagen Molecular (I3M), Centro Mixto CSIC - Universitat Polit\`ecnica de Val\`encia, Camino de Vera s/n, Valencia, E-46022, Spain}
\affiliation[21]{
Laboratorio de F\'isica Nuclear y Astropart\'iculas, Universidad de Zaragoza, Calle Pedro Cerbuna, 12, Zaragoza, E-50009, Spain}

\date{August 30, 2019}

\abstract{
Motivated by the possibility of guiding daughter ions from double beta decay events to single-ion sensors for barium tagging, the NEXT collaboration is developing a program of R\&D to test radio frequency (RF) carpets for ion transport in high pressure xenon gas.  This would require carpet functionality in regimes at higher pressures  than have been previously reported, implying correspondingly larger electrode voltages than in existing systems.  This mode of operation appears plausible for contemporary RF-carpet geometries due to the higher predicted breakdown strength of high pressure xenon relative to low pressure helium, the working medium in most existing RF carpet devices. In this paper we present the first measurements of the high voltage dielectric strength of xenon gas at high pressure and at the relevant RF frequencies for ion transport (in the 10 MHz range), as well as new DC and RF measurements of the dielectric strengths of high pressure argon and helium gases at small gap sizes.  We find breakdown voltages that are compatible with stable RF carpet operation given the gas, pressure, voltage, materials and geometry of interest.}


\begin{document}
\maketitle

\section{Introduction}
\label{sec:intro}

Radio frequency (RF) charged particle manipulation techniques have been
successfully used for several decades to trap and control ions~\cite{dehmelt1962orientation}.  As these techniques are becoming more
efficient, they are being applied to a wider range of applications and
environments.  For example, RF carpets~\cite{wada2003slow} have been shown to
efficiently transport ions in a helium buffer gas for substantial distances in pressures up to
300~mbar~\cite{gehring2016research,ranjan2011new,arai2015performance}. Based on our still preliminary simulations, it appears that they could also be operated efficiently in a xenon buffer gas at
pressures as high as 10~bar. In order to achieve stable motion, the higher
pressure can be compensated for with higher RF voltages and smaller electrode
spacing~\cite{schwarz2011rf}. 

The RF voltage
and electrode spacing in most devices using RF carpets are limited by dielectric breakdown of the buffer gas.
The breakdown voltage is commonly described in terms of a Paschen curve~\cite{paschen1889ueber}, a supposedly universal function of the pressure times the gap distance for each gas.  We expect the breakdown voltage in xenon to be significantly higher than achievable in helium, and with a beneficial scaling with pressure. There is also evidence that Paschen's law is frequency dependent with higher RF frequencies corresponding to lower breakdown voltages~\cite{jones1951high,townsend1958electrical,sato1997breakdown,lisovskiy1998rf}. The RF voltage
breakdown strength in high pressure xenon gas at sub-millimeter distances has not
been previously experimentally measured.

Our study of RF voltages in noble gases is motivated by the search for
neutrinoless double beta decay in $^{136}$Xe.  Observation of neutrinoless double beta decay
($0\nu\beta\beta$) would prove that neutrinos are Majorana fermions and that
lepton number is not conserved, and it could explain the matter-antimatter
imbalance in the universe through leptogenesis~\cite{fukugita174baryogenesis}.
If $0\nu\beta\beta$ does exist, the expected rate would be very small, and
detection will require ton-scale detectors with unprecedented low background rates. 
Since known background processes induce nuclear Z+2 transitions at an insignificant rate~\cite{moe1991detection}, a background free measurement of double beta decay events could be achieved with the identification of the
daughter nucleus.  $^{136}$Xe is a common nuclide in neutrinoless
double beta decay searches, and there are currently several ongoing efforts to
identify the single barium ions that result from their
decay~\cite{sinclair2011prospects,brunner2013setup,twelker2014apparatus,mong2015spectroscopy,brunner2015rf,nygren2016detection,jones2016single,byrnes2019progress,mcdonald2018demonstration,chambers2018imaging,thapa2019barium,byrnes2019barium,rivilla2019towards}.  

For any of these barium tagging proposals to be successful, new technology 
will need to be developed to identify the barium ion  within, or extracted from, the active volume
of the xenon detector. In the case of high pressure gaseous xenon (HPGXe) time projection chambers (TPCs),
used by the NEXT collaboration~\cite{martin2016sensitivity}, the barium ions are expected to be highly ionized initially, but will then
 undergo partial neutralization until they reach the doubly ionized state
(Ba$^{++}$). The ion is slowly transported in various solvation states to the cathode via the drift electric field ~\cite{bainglass2018mobility,Novella2018measurement}.
Transport along the cathode to a small scanning region could be achieved using
RF carpet ion manipulation techniques.  Concepts based on RF have been previously explored in the context of double beta decay by the nEXO collaboration, most notably in~\cite{brunner2015rf}. There, an RF funnel was used to concentrate ions from a fast gas flow into an orifice for extraction to a vacuum ion trap. The RF carpet concept favored by the NEXT collaboration does not involve gas flow, but rather ion transport lateral to the cathode surface, leading the ion to an internal sensor operating in the high pressure gas.

Validity of the concept of RF carpet transport in high pressure xenon relies critically on being able to apply sufficiently high voltages to drive the ions without significant losses.  Preliminary simulation studies~\cite{woodruff2019barium} indicate that we can
transport Ba$^{++}$ in a xenon buffer gas at 10~bar with RF amplitudes as low
as a few hundred volts with sub-millimeter electrode spacings. This would be significantly above what is typically realized in stopper cells, where dielectric breakdown of the low pressure helium buffer gas is limiting voltages above approximately 100V~\cite{postel2000vacuum,sosov2004determination}.

While there are no existing RF breakdown measurements in the pressure ($P$) and gap
distance ($d$) regions of interest, there have been RF and DC measurements near the
 $P\times d$ region, a universal combination under the Paschen formalism. DC breakdown in low pressure argon
gas (up to 100~Torr) was reported in Refs.~\cite{hirsh2012gaseous,lieberman2005principles,farag2013mixing,lisovskiy2000low} in the $P\times d$ range up to 10~Torr~cm
with gap distances of several cm. Systematic studies of the effect of electrode
material and stressed area on DC breakdown in argon in the $P\times d$ region
were carried out in~\cite{farag2013mixing}. RF ($f=13.56$~MHz) breakdown in low
pressure argon (up to 20~Torr) was measured in
Refs.~\cite{lisovskiy1998rf,smith2003breakdown,sato1997breakdown} up to
30~Torr~cm also with gap distances of several cm. In low pressure xenon (up to 500~Torr),
measurements of the DC breakdown voltage in $P\times d$ range including 0.3 to
12~Torr~cm were reported in
Refs.~\cite{bhattacharya1976measurement,jacques1986experimental,postel2000vacuum}, and the
voltage breakdown at microwave frequencies ($f=2800$~MHz) covering the same
$P\times d$ range was measured in Ref.~\cite{bradford1959electrical}. In
low pressure helium (up to 600~Torr), voltage breakdown was measured covering
the same $P\times d$ range with gap distances as small as 1~mm at RF frequencies ($f=13.56$~MHz) in Refs.~\cite{park2001gas,moravej2004physics} and with DC in Ref.~\cite{borg1994investigation}.
 
In this paper, we describe our measurements of the Paschen curve for DC and RF
voltages in the pressure and electrode gap ranges of interest for operating RF
carpets in high pressure gases. In addition to xenon, we measured the breakdown
voltages in argon and helium (which are more commonly used as RF carpet buffer
gases) in pressures from 100~mbar to 10 bar. We show in argon that tripling the
electrode gap distance from 0.005~in. (0.13~mm) to 0.015~in. (0.38~mm) has no
significant effect on the breakdown voltage at fixed $P \times d$, so we only include the 0.005~in. gap
in the helium and xenon measurements. All of the RF measurements are performed
at a frequency of 13.56~MHz. It has been shown in
Ref.~\cite{pang2011development} that in the low MHz frequency regime, the
efficiency of RF carpet transport increases with increasing frequency up to
approximately 10~MHz at which point the efficiency stays constant, and so 13.56~MHz is chosen to be a convenient frequency within our range of interest.

In Sec.~\ref{sec:experiment} we describe the experiment, including the gas
handling system in Sec.~\ref{sec:gashandling}, the
electrode configuration in Sec.~\ref{sec:electrode}, the DC and RF delivery and
monitoring systems in Sec.~\ref{sec:rfdelivery}, and the data taking methodology
including our estimations of systematic uncertainties in Sec.~\ref{sec:systematics}.

\section{Experiment}
\label{sec:experiment}

\begin{figure}
    \begin{centering}
    \includegraphics[width=1.0\columnwidth]{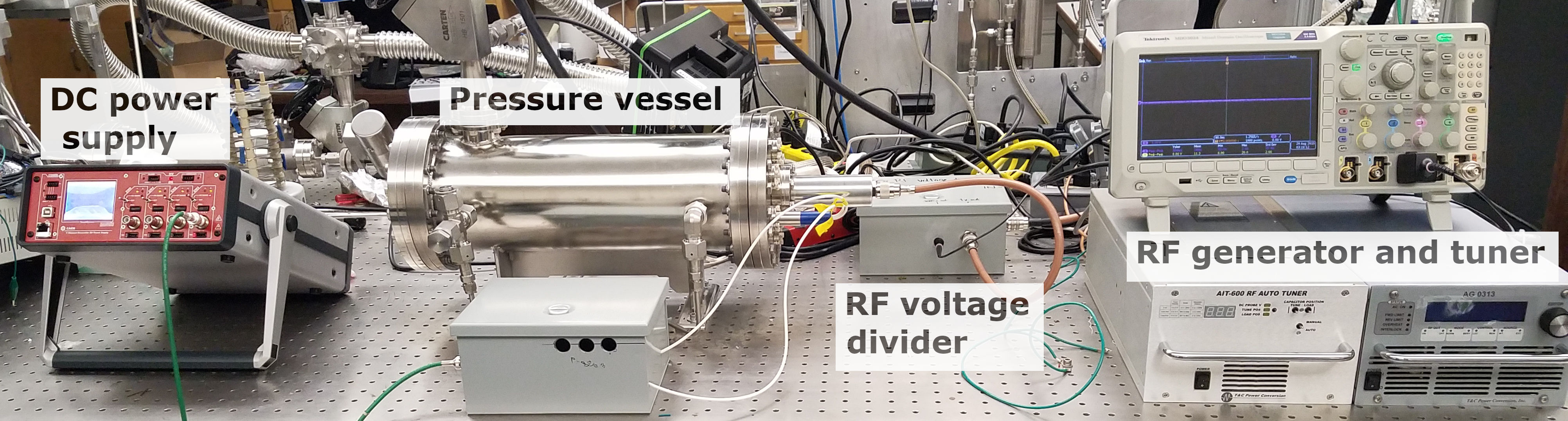}
    \par\end{centering}
    \caption{Photograph of the experimental setup. \label{fig:lablabels}}
  \end{figure}

    The experimental setup, shown in Fig.~\ref{fig:lablabels}, consists of two
electrodes connected to (RF or DC) high voltage (HV) inside of a pressure
vessel capable of holding pressures up to 10~bar. For each gas, three breakdown measurements are taken per pressure. When the vessel is at the desired gas pressure, the RF or DC high voltage is increased until breakdown is observed. For the RF breakdown measurements, the voltage is monitored through an oscilloscope, described in Sec.~\ref{sec:rfdelivery}. When breakdown occurs, the RF circuit goes out of resonance, and the signal on the oscilloscope collapses. For the DC breakdown measurements, the voltage is monitored on the DC power supply itself. When DC breakdown occurs, the voltage drops and the current increases above the set operating current. Any change in the measured DC current from the set value was considered a voltage breakdown. For both RF and DC, the highest voltage observed before the breakdown is recorded as the breakdown voltage.
  
\subsection{Gas System}\label{sec:gashandling}

The electrodes are housed inside a five-liter pressure vessel connected to
a gas handling and circulation system. The same gas system was used in previous studies described in Refs.~\cite{mcdonald2019electron,rogers2018high}, where further information can be found.  The gases used are ultra-high purity (99.999\%)
grade helium and argon from AirGas, and xenon from Concorde Specialty Gases.
Before each gas fill, the system is
evacuated to better than 10$^{-5}$ mbar pressure. The xenon is filled through a
MonoTorr PS-4-MT3 hot getter to the maximum pressure and circulated using a
MetroCad PumpWorks PW-2070 pump with flow rates between one and ten liters per
minute until the pressure vessel volume has been circulated through the getter
ten times. The process is the same for the helium and argon gases, except they
are filled and circulated through a Saes PureGas MicroTorr HP190-902-FV cold
getter instead of the hot getter. HV breakdown measurements are taken at the
highest pressure, then at discrete steps down in pressure as the gas is either
vented into a separate evacuated vessel (in the case of helium and argon) or reclaimed (in the case of xenon). The gas pressure in the vessel is monitored with an Omega pressure transducer with 10
mbar precision. The measurements were taken at pressures from 0.1 to 10~bar.
The maximum breakdown voltage achievable was 500~V, discussed in Sec~\ref{sec:results}, limiting the maximum measurement pressures to 1~bar in xenon and 4~bar in argon. The minimum pressures measurable in argon and helium are limited by the discrete (1~watt) step sizes of the RF voltage generator, creating a lower pressure measurement limit of 0.2~bar in argon and 1~bar in helium.

  \subsection{Electrode Configuration}\label{sec:electrode}

  \begin{figure}
    \begin{centering}
    \includegraphics[width=1.0\columnwidth]{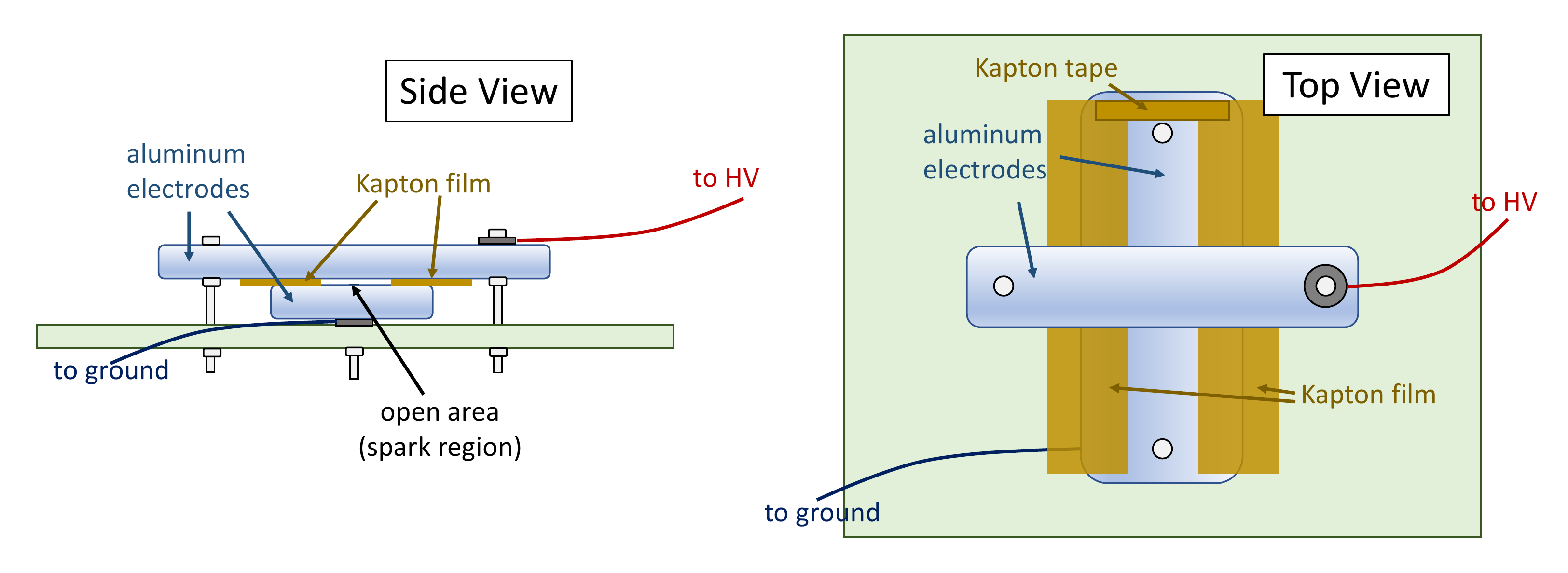}
    \par\end{centering}
    \caption{Diagram of the electrode configuration. \label{fig:electrodediagram}}
  \end{figure}

    The electrodes consist of two aluminum bars positioned in a cross, as shown
in Fig.~\ref{fig:electrodediagram}. The top bar is 2.5~in. (64~mm) long by 1.0~in. (25~mm)
wide, and the bottom bar is 2.5~in. (64~mm) long by 0.5~in. (13~mm) wide. 
The cross
configuration is used to reduce the amount of electrically stressed area, which has been shown to affect the breakdown voltage~\cite{okawa1988area}. The top bar is
connected to the HV signal, and the bottom bar is connected to ground. Two
pieces of thin, 0.005~in. (0.13~mm) Kapton film, with 0.0005~in. thickness tolerance, placed 0.4~in. (10~mm) apart, are sandwiched in
between the electrodes to set the gap distance.
The voltage breakdown generally happens in the open area between the electrodes where no
Kapton is placed. The position of the breakdown can be determined from traces left on the electrode surface. The total open (stressed) area between the electrodes
is 0.2~in.$^2$ (130~mm$^2$). The entire apparatus is held in place with nylon screws onto 
a large (3~in.~$\times$~4~in. or 75~mm~$\times$~100~mm) piece of HDPE.

  \subsection{HV Delivery and Monitoring System}\label{sec:rfdelivery}
  \begin{figure}
    \begin{centering}
    \includegraphics[width=0.85\columnwidth]{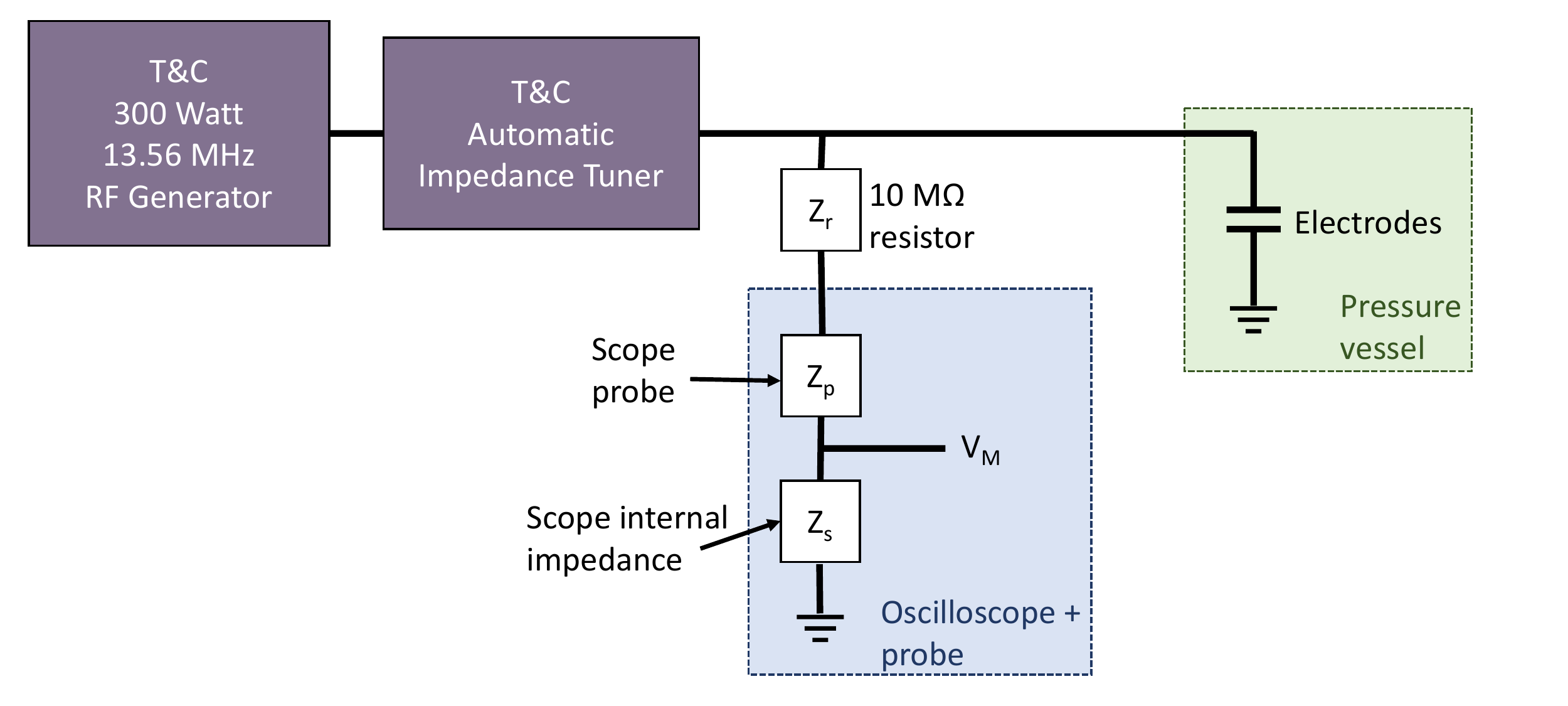}
    \par\end{centering}
    \caption{Diagram of the RF circuit. \label{fig:rfcircuit}}
  \end{figure}

\begin{figure}
    \begin{centering}
    \includegraphics[width=0.45\columnwidth]{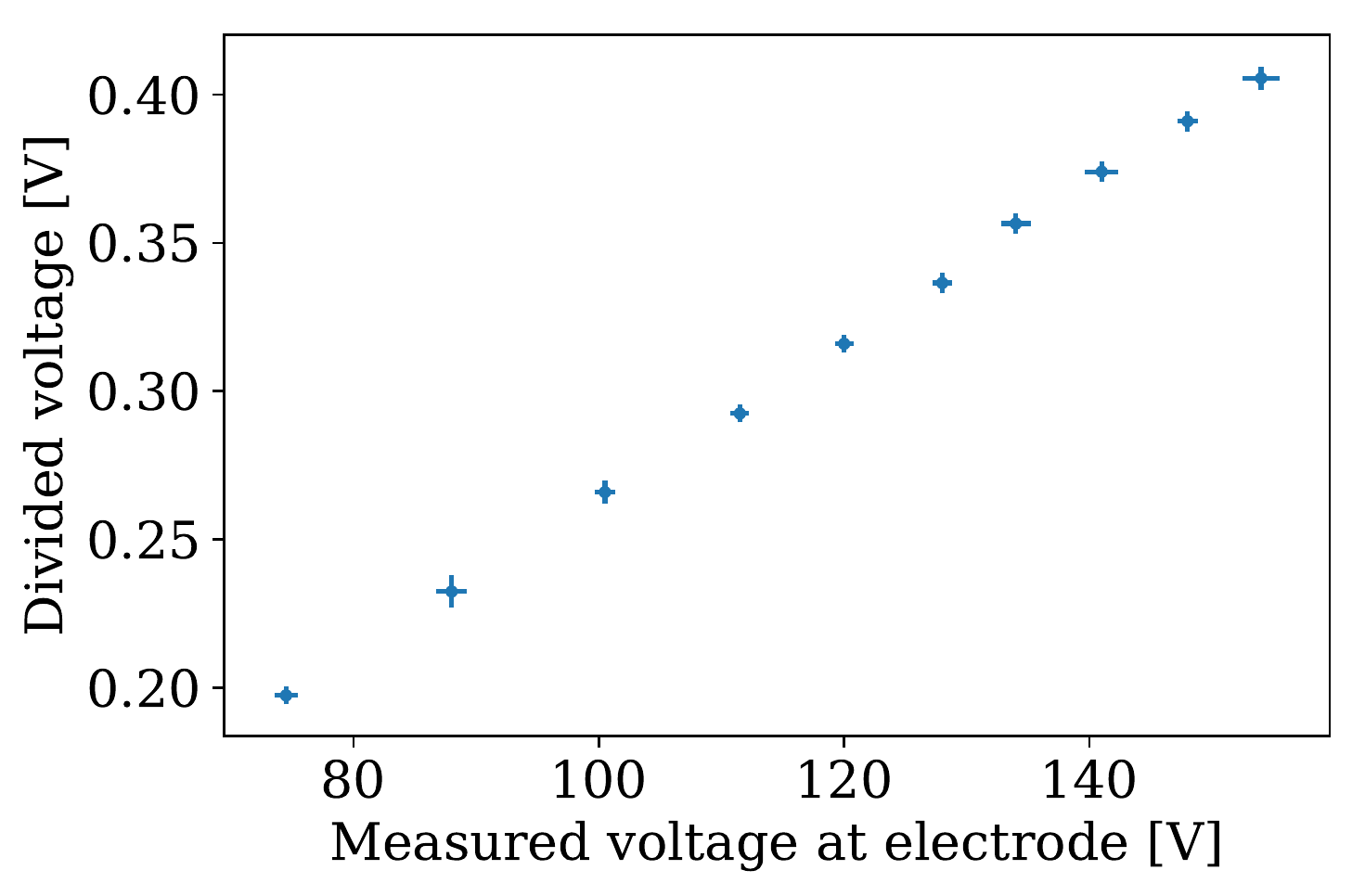}
    \includegraphics[width=0.45\columnwidth]{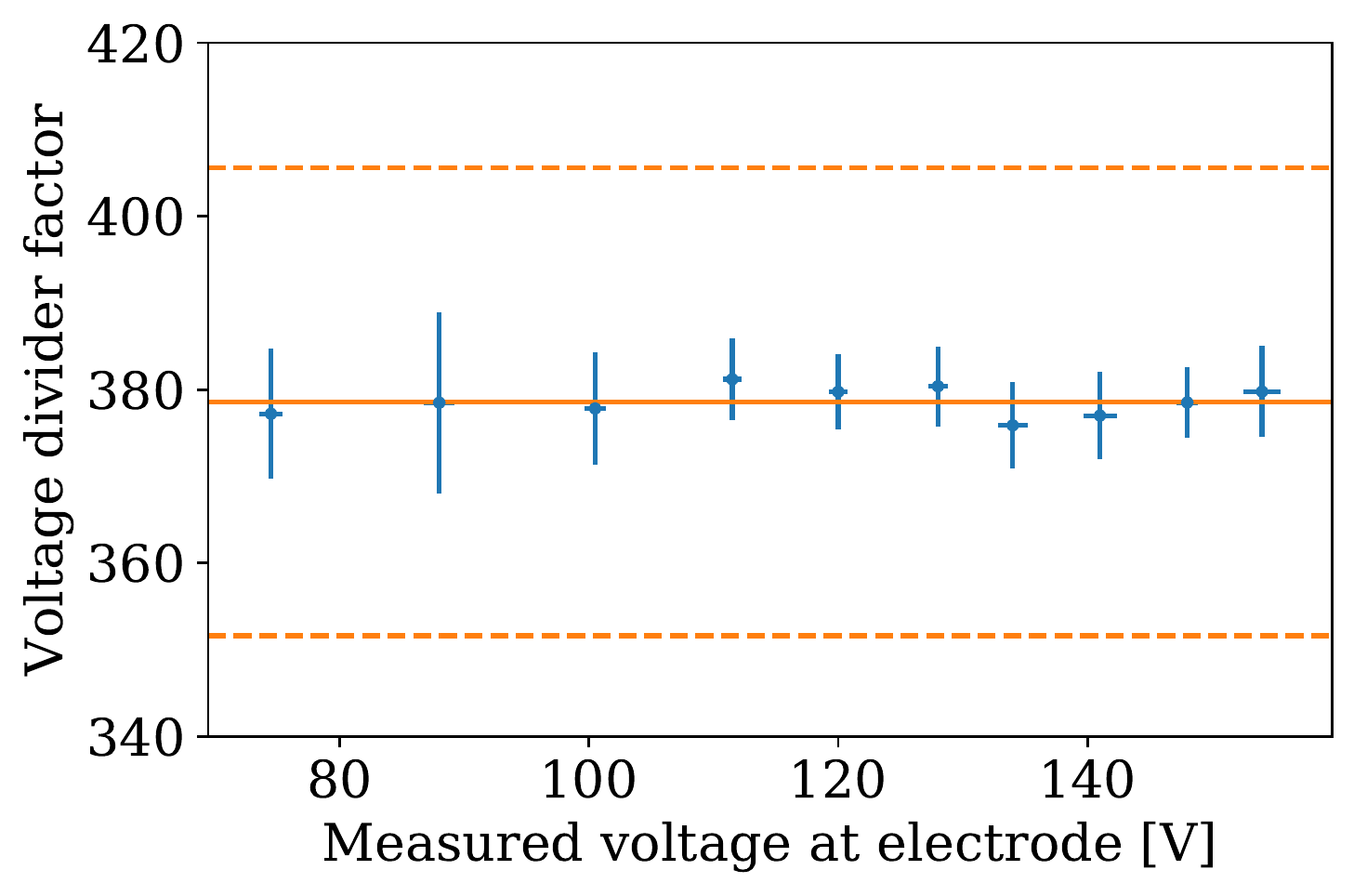}
    \par\end{centering}
    \caption{Example voltage divider measurement. The left plot shows the
    comparison of the voltage amplitude measured at the electrode and the
    voltage amplitude measured after the potential divider on an oscilloscope with the uncertainty on the voltage measurement. The right plot shows
    the unitless ratio of the two. The error bars on the points show the uncertainty on the peak voltages measured by the oscilloscope at the electrode and after the potential divider. The solid line in the right plot shows the average
    voltage divider factor, and the dashed lines show the combined (one standard deviation) uncertainty on the
    true voltage divider factor, discussed in Sec~\ref{sec:systematics}. \label{fig:vdivide}}
  \end{figure}

    The RF HV is supplied by a 13.56~MHz, 300-watt AG Plasma Series generator
and an automatic impedance tuner from T\&C Power Conversion. The RF voltage is
actively monitored through a potential divider to an oscilloscope, as shown in
Fig.~\ref{fig:rfcircuit}. The potential divider consists of a 10~M$\Omega$
resistor (labeled $Z_r$), the oscilloscope probe (labeled $Z_p$), and the
internal impedance of the oscilloscope itself (labeled $Z_s$). The impedance of
each component at 13.56~MHz is not precisely known, but the voltage division
factor can be measured. We found that the exact voltage division factor varied
with different electrode spacings and impedance tuner settings, presumably due to changing capacitances and inductances within the circuit impacting the impedance at radio frequencies. Therefore, it was
measured directly for each configuration. Fig.~\ref{fig:vdivide} shows an
example measurement of the voltage divider factor for one electrode
configuration. The left plot shows the voltage amplitude measured after the
potential divider as a function of the voltage amplitude measured at the HV
electrode, and the right plot shows the voltage divider factor with the total
uncertainty on the conversion to true voltage at the HV electrode, discussed in Sec~\ref{sec:systematics}. The RF
signal is delivered to the electrodes inside a pressure chamber through a
CeramTec copper RF power feedthrough.  The DC HV is supplied by a CAEN
DT1471HET desktop HV power supply.

  \subsection{Estimation of Systematic Uncertainty}\label{sec:systematics}

    The sources of systematic uncertainty that we took into account are the
uncertainty in the gas pressure measurement, the uncertainty in the spacing
between the electrodes, and the uncertainty in the voltage measurement. We also
tested the effect on the breakdown voltage due to the electrode material and
the size of the stressed area.

The uncertainty in the gas pressure measurement is 10 mbar and is due to the
precision of the transducer. The uncertainty on the gap distance is due to the
compression of the Kapton film between the electrodes and the flatness of the
electrodes. Based on the large elastic modulus of Kapton and absence of any
observable electrode curvature when compared to a granite surface plate, the
uncertainty in the gap distance is estimated to be less than 10\%.

The uncertainty in the RF breakdown voltage is due to the measurement of the
voltage at the electrode and to irreducible error from the randomness in the
breakdown process itself. The uncertainty in the RF measurement is estimated from
the variance in the voltage division factor, described in
Sec.~\ref{sec:rfdelivery}, and in the change in the voltage measured at the HV
electrode using different, compensated, oscilloscope probes, which is found to
be 5\%. To account for the irreducible error in both the RF and DC breakdown voltages, three measurements were taken
consecutively at each pressure, and the standard deviation of the measured
breakdown voltages was added in quadrature to the uncertainty on the voltage
measurement. The gas was not recirculated between the consecutive measurements at each pressure, but no trend was observed between the breakdown voltage and the order of the measurements.

\section{Results and Discussion}
\label{sec:results}
\begin{figure}
    \begin{centering}
    \includegraphics[width=1.0\columnwidth]{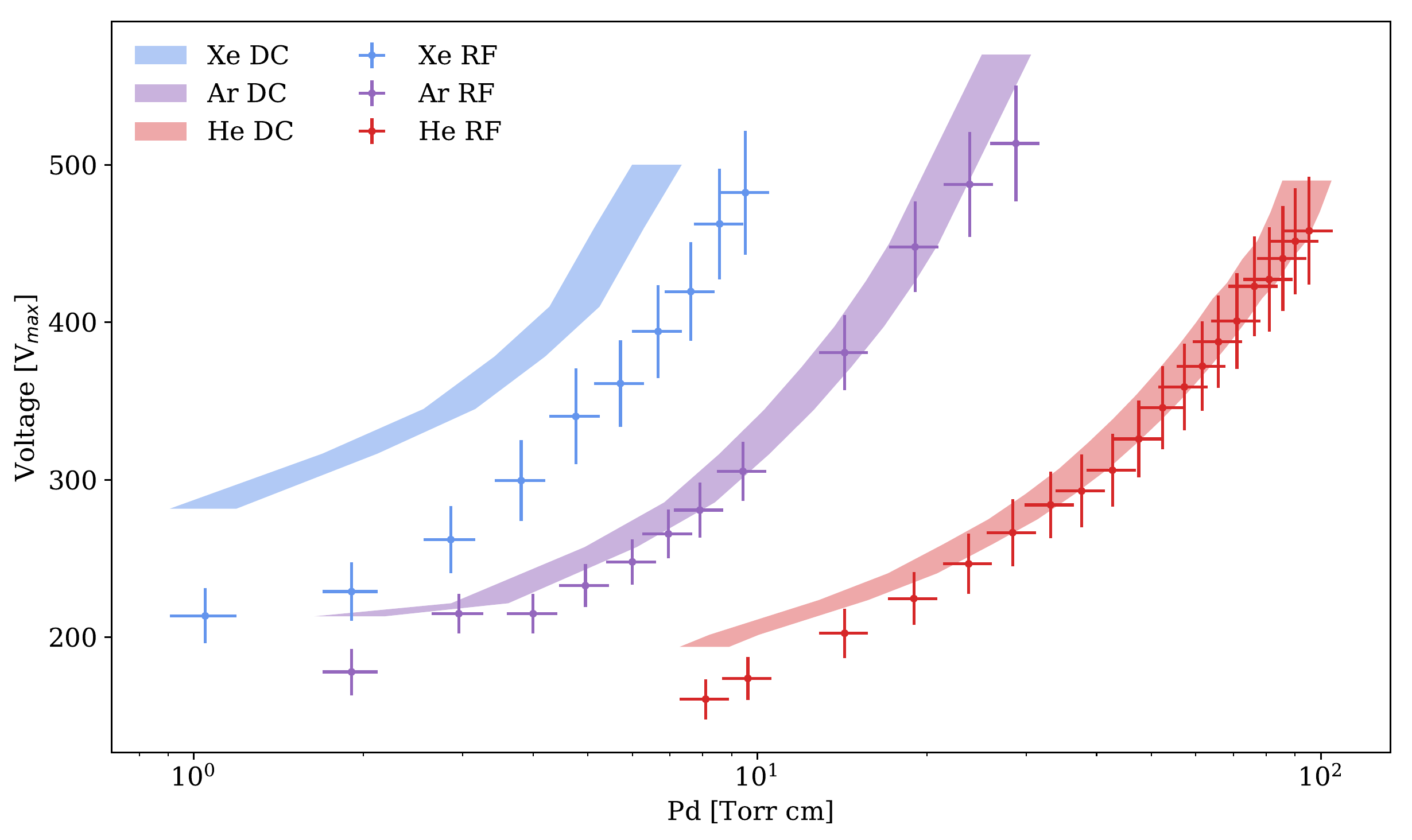}
    \par\end{centering}
    \caption{RF and DC breakdown voltages in xenon, argon, and helium using
aluminum electrodes with a 0.005~in. (0.13~mm) gap. The RF voltage shown is the maximum
amplitude of the RF signal with the total estimated uncertainty. The bands cover the DC breakdown data points including the total estimated uncertainty. The RF and DC data points were taken at the same pressures. \label{fig:rfbreakdowns}}
  \end{figure}

Fig.~\ref{fig:rfbreakdowns} shows the measured RF and DC breakdown voltages
in helium, argon, and xenon with a 0.005~in. gap distance. In the case of both xenon and argon gases, breakdown through the Kapton spacer ultimately became limiting at a field strength of 40~kV/cm.  Before breakdown through the Kapton became prevalent, inspection of traces on the electrode surfaces showed breakdown was across the gas gap, rather than along the insulating surface.  All gases show the expected Paschen-like increase of HV strengh with increasing pressure under both RF and DC.  The RF breakdown voltages are found to be systematically lower than the DC ones, an effect which is most significant in the xenon gas measurements, and represents an approximately 30\% effect. The dependence of the Paschen curve on both $P\times d$ and $f \times d$, where $f$ is frequency, was first shown in Refs.~\cite{jones1951high,townsend1958electrical} and can be quantified if the electron ionization rate and mobility is well known for the given gas and pressure~\cite{kihara1952mathematical,sato1997breakdown,lisovskiy1998rf}. The dependence on frequency decreases with large $P\times d$~\cite{lisovskiy1998rf}. This would explain the larger effect seen in the xenon measurements which are near the Paschen minimum.

\begin{figure}
    \begin{centering}
    \includegraphics[width=1.0\columnwidth]{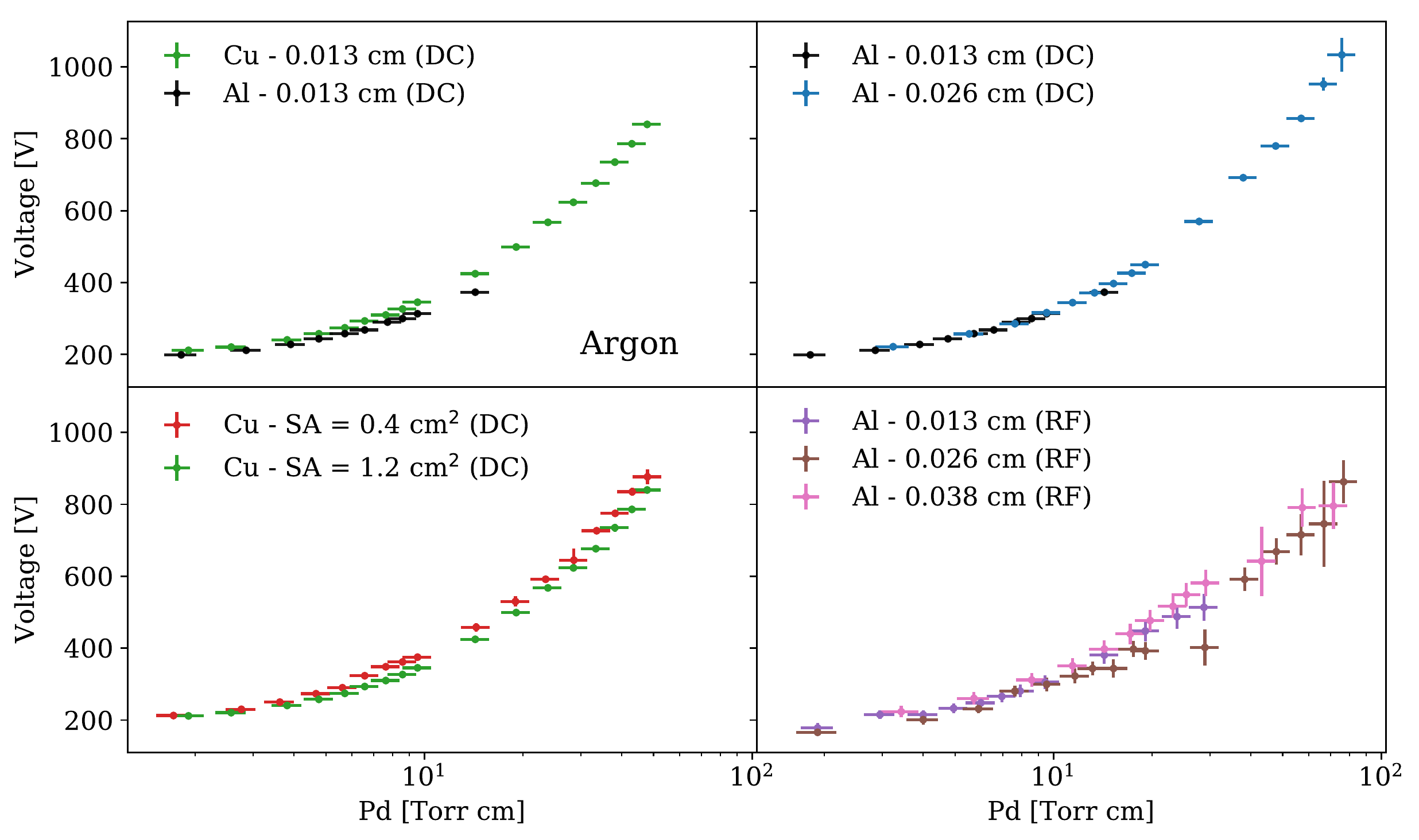}
    \par\end{centering}
    \caption{Comparison of breakdown voltages in argon using different
electrode configurations. The top left plot shows the ($\sim$8\%) effect of using two different
electrode materials, the bottom left plot shows the ($\sim$6\%) effect of two different
stressed areas, the right plots show the effect of different gap distances
with DC on the top and RF on the bottom.  \label{fig:breakdownsyst}}
  \end{figure}

When designing this test, we investigated several possible variables that may correlate with breakdown strength, with results shown in Fig.~\ref{fig:breakdownsyst}.  To determine whether electrode material is significantly correlated with breakdown strength, the breakdown voltage in argon was measured with copper as well as aluminum electrodes.  The work functions of these metals are sufficiently different (4.53--5.10~eV for Cu compared to 4.06--4.26~eV for Al~\cite{holzl2006solid}) that any substantial dependence through photoelectron emission would likely be revealed.  Fig.~\ref{fig:breakdownsyst} shows that there is some systematic trend, but it is rather small, barely larger than systematic uncertainty.  We tested dependence on stressed area using similar 
electrodes with a larger Kapton insulator resulting in 1/3 of the original electrode surface available for breakdown.
The results of these tests are shown in Fig.~\ref{fig:breakdownsyst} lower left. Again, a small effect is observed, but it appears consistent with the scale of systematic uncertainty.  We also tested Paschen scaling by changing the gap size. We observed that changing the gap distance and pressure had no significant effect on either the RF of DC breakdown
voltages at fixed $P\times d$, validating  Paschen-like behaviour in this regime.  These comparisons are shown for both RF and DC voltages in the right panels of Fig.~\ref{fig:breakdownsyst}.

\subsection{Comparison with Published Data}

\begin{figure}
    \begin{centering}
    \includegraphics[width=0.49\columnwidth]{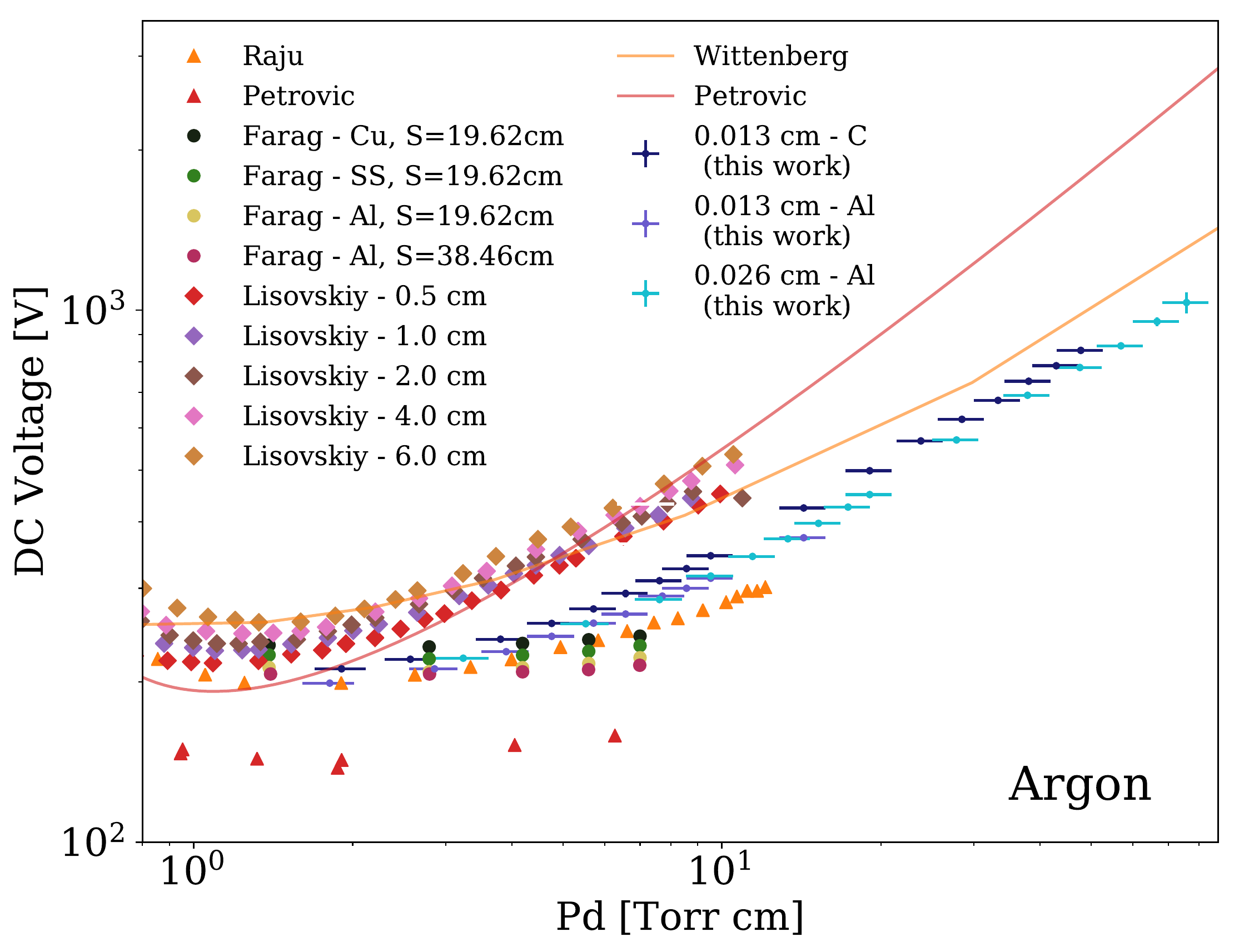}
    \includegraphics[width=0.49\columnwidth]{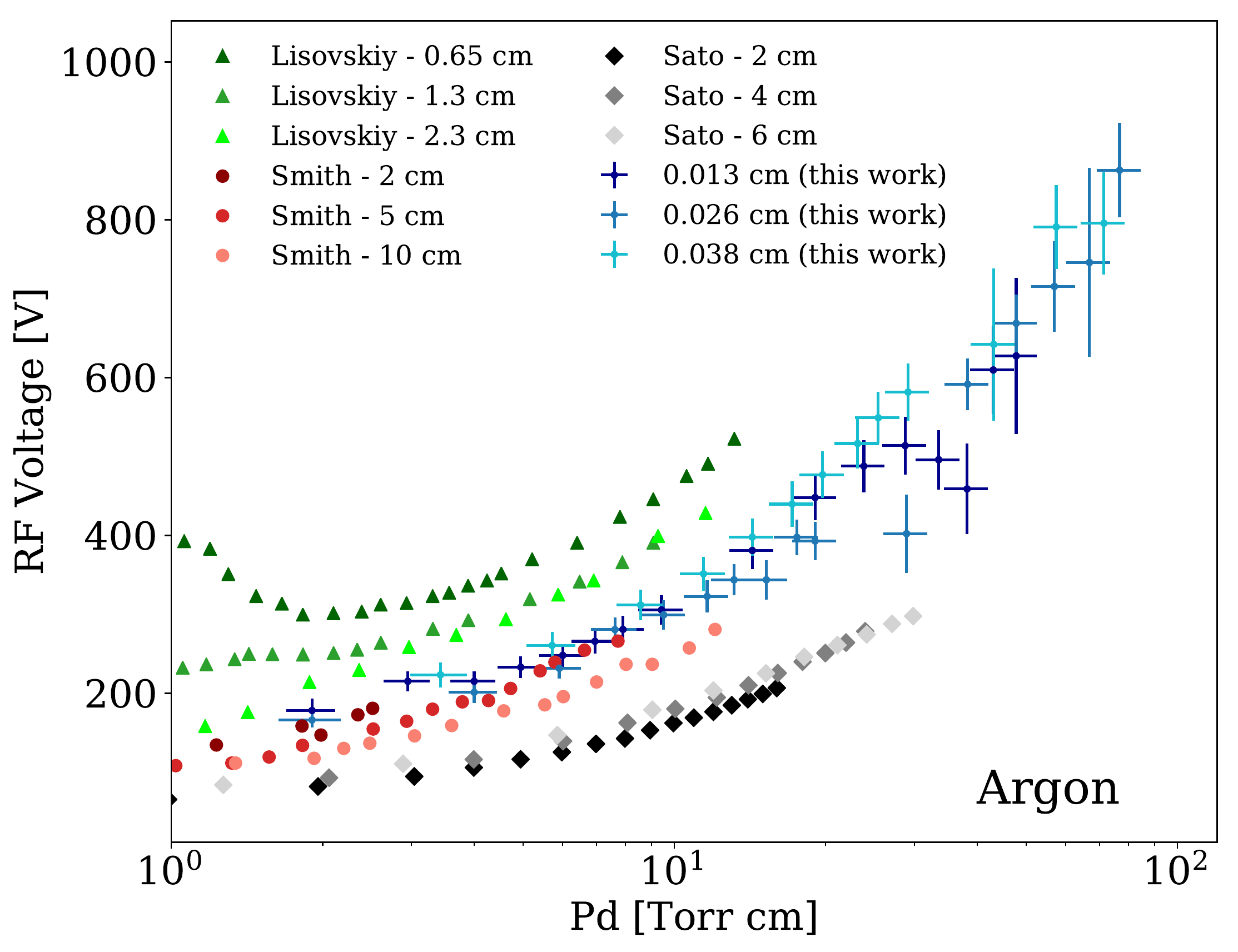}
    \par\end{centering}
    \caption{Comparison of DC (left) and RF (right) voltage breakdown in argon
gas to a selection of published data. The RF measurements were all taken at
13.56~MHz, and the maximum voltage amplitude is reported. In the left plot, the
triangles are reported in
textbooks~\cite{lieberman2005principles}~and~\cite{hirsh2012gaseous}, the
circles~\cite{farag2013mixing} represent different electrode materials (Cu, Al,
and stainless steel) and different stressed areas (19.62~cm and 38.46~cm) with Cu
electrodes. The diamonds were measured in Ref.~\cite{lisovskiy2000low}. The lines are the parameterized curves in Refs.~\cite{wittenberg1962gas,lieberman2005principles}. In the
right plot, the triangles were measured in Ref.~\cite{lisovskiy1998rf}, the
circles in Ref.~\cite{smith2003breakdown}, and the diamonds in
Ref.~\cite{sato1997breakdown}. \label{fig:argoncompare}}
  \end{figure}

Fig.~\ref{fig:argoncompare} shows the comparisons of the DC (left) and RF
(right) breakdown voltages in high pressure argon to a selection of published
measurements in low pressure argon in the same $P\times d$ range. In both DC
and RF, the measurements fall in the middle of the spread of published data. The
behavior of the comparison between the different electrode materials and
stressed areas from Farag \textit{et al.} in Fig.~\ref{fig:argoncompare} left
agree well with the behavior shown in Fig.~\ref{fig:breakdownsyst} on the lower left.

\begin{figure}
    \begin{centering}
    \includegraphics[width=0.49\columnwidth]{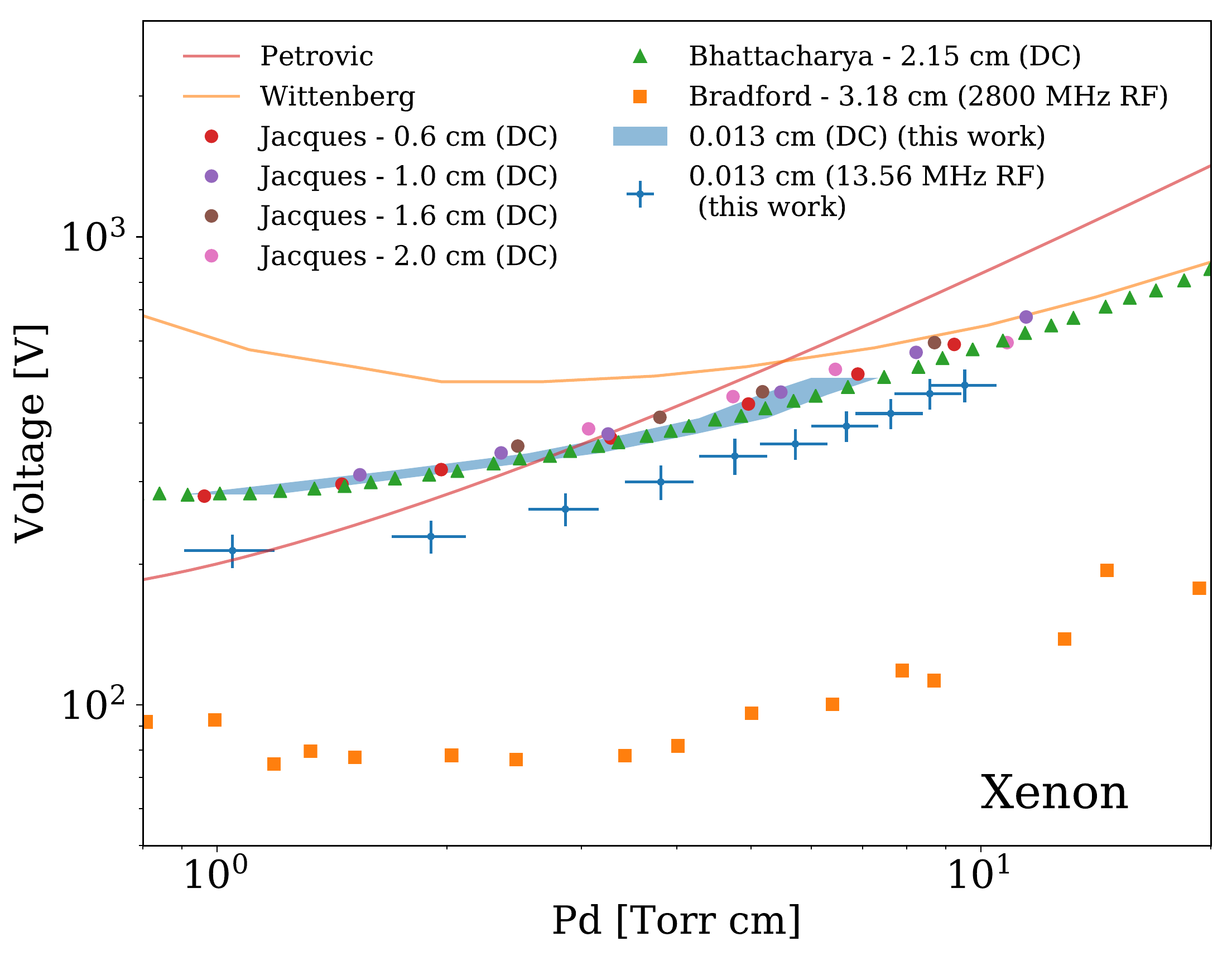}
    \includegraphics[width=0.49\columnwidth]{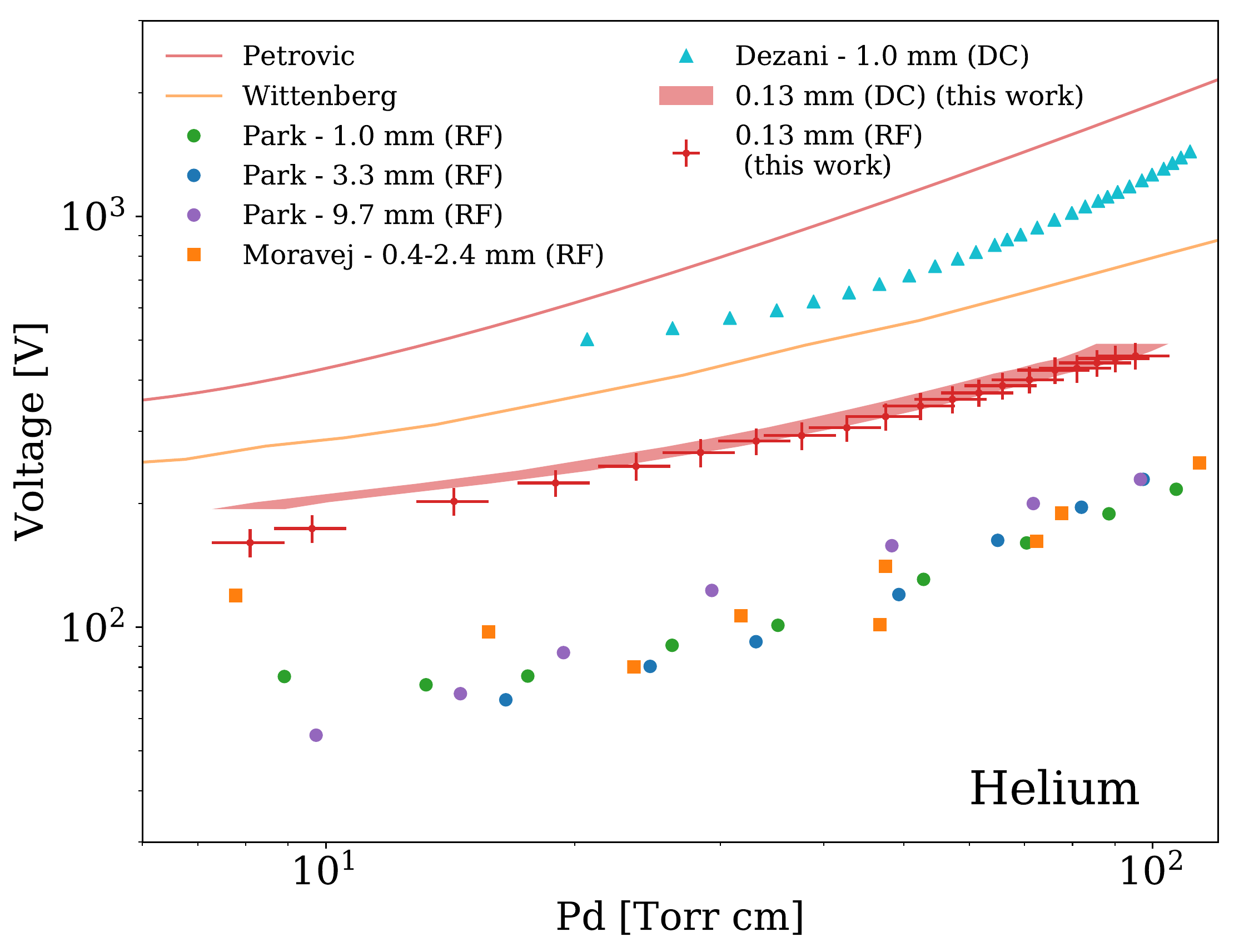}
    \par\end{centering}
    \caption{Comparison of xenon (left) and helium (right) RF and DC voltage
breakdown measurements to a selection of published data. The DC measurements in
this work are shown as shaded area in both plots, and the RF measurements
are shown as blue points with error bars for xenon and red points with
error bars for helium. In the left plot, the DC circles~\cite{jacques1986experimental}
were measured using gold electrodes, the DC triangles~\cite{bhattacharya1976measurement}
were measured using nickel electrodes, and the squares~\cite{bradford1959electrical}
represent microwave breakdown measurements at 2800~MHz.  In the right plot, the
circles~\cite{park2001gas} were measured with aluminum electrodes, and the
squares~\cite{moravej2004physics} were measured with several gap distances from
0.4--2.4~mm, and the DC triangles~\cite{borg1994investigation} were measured using an aluminum electrode. The lines are the parameterized curves in Refs.~\cite{wittenberg1962gas,lieberman2005principles}.
\label{fig:xenoncompare}}
  \end{figure}

    Fig.~\ref{fig:xenoncompare} shows the comparisons of the DC and RF
breakdowns in xenon (left) and helium (right) to a selection of published data.
The previous published data were taken with gap distances in the range of 6--32~mm
for xenon and 0.4--10~mm for helium at sub-atmospheric pressures compared to ours,
taken at a 0.005~in. (0.13~mm) gap distance.  The DC breakdown measurements
(shaded area) in xenon agree with published data, and the RF breakdown
measurements fall between the published DC and microwave breakdown
measurements. Both the helium RF and DC measurements in
Fig.~\ref{fig:xenoncompare} right fall well above the published RF measurements,
also taken at 13.56~MHz, and below the the published DC measurements. 

\begin{figure}
    \begin{centering}
    \includegraphics[width=0.85\columnwidth]{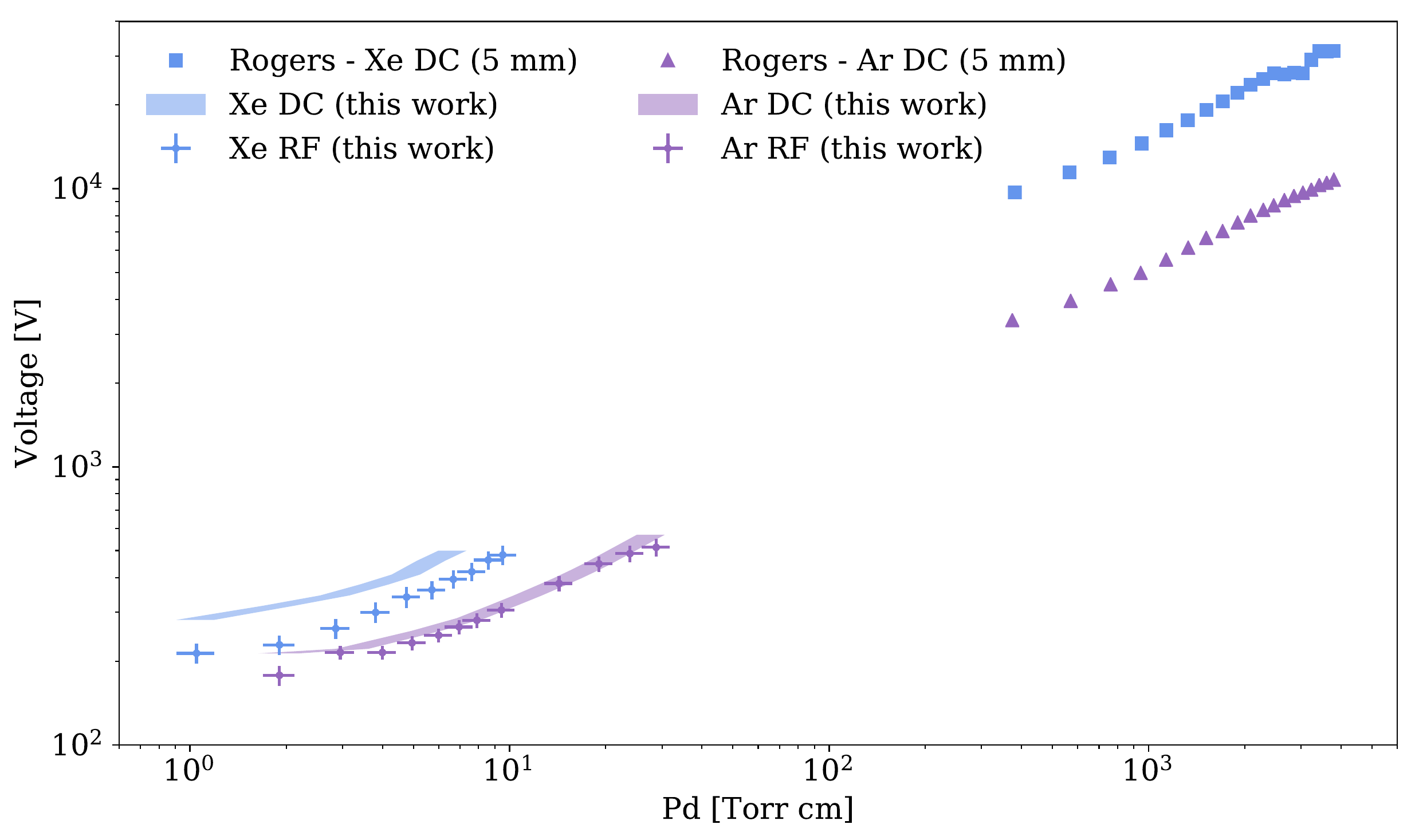}
    \par\end{centering}
    \caption{Xenon and argon RF and DC voltage breakdown measurements from this work with previous high pressure xenon and argon DC breakdown measurements with larger gap distance from Ref.~\cite{rogers2018high}.
  \label{fig:rogerscompare}}
  \end{figure}

    Fig.~\ref{fig:rogerscompare} shows the agreement between our DC and RF
breakdown measurements in argon and xenon and high pressure, large gap-size DC breakdown
measurements performed previously by this group described in
Ref.~\cite{rogers2018high}. The previous measurements were taken at similar pressures within the same gas system, with gap distances of 5~mm.
  
\subsection{Implications for RF Carpets in High Pressure Xenon}

The pressure ranges where in-gas breakdowns could be studied were 1 to 10~bar in helium, 0.2 to 4~bar in argon, and 0.1 to 1 bar in xenon.  Since RF carpets with small electrode pitches are commonly printed on a Kapton substrate, our data strongly suggest that the Kapton will be the limiting factor in their operation in xenon at 10 bar, rather than through-gas breakdowns, as are limiting in the case of low pressure helium. For a 0.005~in. pitch carpet, the threshold of operation may thus be close to 500~V peak-to-peak.  It is notable, however, that in such a carpet the copper electrode
traces are laid on top of the Kapton and not on either side as in the
configuration here.  This implies a smaller electric field through the
Kapton itself, and thus our measurements here may be considered as a cautious lower limit. Future voltage
breakdown tests of printed copper electrode traces on a ``flex'' Kapton board
will determine if this is true. 

Comparing this experimental constraint with our simulations, we observe that significant parameter space remains open where RF carpet operation is theoretically predicted to be possible in xenon gas at 10 bar.  Preliminary simulations have shown that Ba$^{++}$ ion transport in xenon buffer gas at 10 bar should be achievable with RF voltages as low as 250~V, given existing electrode pitch geometries with 0.1~mm gap distances~\cite{hamaker2016experimental}. 

The next phase of our research program will be to demonstrate ion transport using a 5~cm scale RF carpet in xenon gas.  The results of this test will help establish the viability of RF-carpet based ion guiding schemes for barium tagging within high pressure xenon gas detectors.

\section{Conclusions}
\label{sec:conclusion}

We have presented the first RF breakdown strength measurements in xenon gas at pressures above 100 Torr and at frequencies in the MHz range, as well as new measurements of RF and DC breakdown voltages in argon at pressures of 100~mbar--1~bar and helium at pressures of 1~bar--10~bar and sub-millimeter gap spacings.

We find that the RF breakdown voltages in high pressure helium, argon, and
xenon for sub-millimeter gap distances agree approximately with existing low pressure data in
the same $P\times d$ range with much larger gap distances. In the cases of
argon and helium in the $P\times d$ region between 10 and 100 Torr cm, the RF and
DC breakdown voltages agree well with each other. There are some deviations
between the xenon RF and DC breakdown voltages, but these measurements all
exist below 10 Torr cm, where we start to observe deviations in argon and helium
as well. Based on our data, the limiting factor in the application to high pressure RF carpets in xenon is likely to be the Kapton substrate that starts to fail at approximately 500~V at
0.005~in. thickness, rather than the gas itself.  This is distinct to the case of helium-based systems, where through-gas breakdowns are limiting. This warrants future tests of breakdown between electrodes
printed on a Kapton substrate.  

These results suggest promise for the concept of using RF carpets in high pressure xenon gas.  Such a device may prove to be a critical ingredient in a workable scheme for barium ion identification within xenon gas time projection chambers.

\bibliography{rfhv}

\end{document}